\documentclass[oneside,a4paper]{article}

\usepackage{hyperref}
\usepackage{authblk}
\usepackage{amsfonts}
\providecommand{\keywords}[1]{\textbf{\textit{Keywords:}} #1}

\title{Counting stars: a survey on flexible Skyline Query approaches.}
\author{Alessandro Del Giudice}
\affil{Politecnico di Milano\\
Milan, Italy\\
\href{mailto:alessandro.delgiudice@mail.polimi.it}{alessandro.delgiudice@mail.polimi.it} }
\date{}

\begin{document}
\maketitle
\begin{abstract} 
Nowadays, as the quantity of data to process began to rise, so did the need for a method to discern what pieces of information could be useful for the user; in response, researchers focused their efforts on improving the already existing ranking methods or creating new ones starting from them. 
This survey will be presented a small list of some of the most known and/or most recent solutions proposed, with some possible applications for them, concerning a state of the art restricted to around the last ten years, comparing their performance with the traditional one top-k and skyline queries.
\end{abstract}

\section{Introduction}
In the last few decades, we have witnessed the constant evolution of the concept of data: its shape changed from a restricted amount of focused information to explore, to much larger pools, bearing the need for filters to discern useful or important data in the context of search. For this purpose, a large number of variations to the already existing ranking queries were proposed, all bearing their strengths and weaknesses. Before exploring the new solutions, the next sections will provide a brief introduction to Top-k and Skylines ranking queries.

\subsection{Top-k approach}\label{topk}
The use of ranking queries is often strictly tied to the need of retrieving the most relevant k entries (\textbf{Top-k}) in a given dataset, considering certain factors that will be established based on the context of use \cite{topKQueryIlyas} and most of the applications focused on retrieving these entries involve a high usage of joining and aggregating functions to compute an acceptable result. A common way that has been developed to extrapolate them more simply involves the usage of a \textit{scoring function}, which returns a \textit{score} given every single item that needs to be analyzed, to allow the successive selection of the top values.
To get an idea of how crucial can be efficiency and precision in such queries, one could think about the amount of data that an online estate agency has to handle daily and how many requests its website will have to address to return interesting results to a certain user.

The major drawback of this approach resides in the low generality of the results prompted, which will be highly dependent only on the factors considered in the scoring function \cite{ORD, restrictedMarti}.


\subsection{Skyline approach}
Skyline ranking queries are relatively recent compared to Top-k, and are entirely based on the mathematical concept of \textit{dominance relationship} or \textit{Pareto dominance}: given such a relation, the Skyline query will return all the objects that cannot be dominated by others \cite{introSkyline}. To have it simplistically, the objects belonging to the skyline won't be worse than any other entry in any dimension of study considered (i.e. they'll be non-dominated tuples) and will be incomparable\footnote{Given two tuples $\textit{s,t}\in \textit{D}$ domain of interest, \textit{s} and \textit{t} are incomparable if simultaneously dominated by and dominating each other.} with each other.\cite{introSkyline, topKDominatingZhu}.

In multidimensional datasets, the definition of skyline query is equivalent to the solution of a known maximum vector problem and this concept led to the development of different algorithms aimed at making the ranking process more efficient, while the interest for the topic itself began to spread more and more. In early works computation of such values was a problem of an algorithmic sort, considering how data was stored in its entirety in main memory, nowadays algorithms for skyline query processing can be divided into index and non-index based. 
Practical uses of this notion can be typically seen on data objects relatable to points in a given Cartesian plane, like the case considered in \ref{topk}: if one wanted to filter by price and distance from their workplace, a glance at the skyline returned would be enough to know which offers can be considered useful for them.

More details on the topic can be found in \cite{introSkyline, kMostLin, kalyvas2017survey}. 

\section{Flexible/Restricted Skylines}
\subsection{Restricted Skyline, ND and PO}\label{res}
A possible solution to the multi-objective optimization problem is proposed in \cite{restrictedMarti}: the framework introduced to combine the generality of skyline queries with the cheaper computation of the top-k approach defines the notion of \textit{Restricted Skyline queries} (R-skylines). Similarly to the \textit{prioritized skyline} introduced in \cite{mindolin}, R-skylines take into account different weights that the attributes can have on determining the best objects to consider, but instead of creating a strict hierarchy between them, arbitrary constraints are defined to allow more flexibility in modeling.\\ \\
Core to this method is the concept of \textit{F}-dominance formalized by the authors, which states that given a set of monotone scoring functions \textit{F} and considering a pair of tuples \textit{t} and \textit{s}, we can define: 

\begin{center}\large{\textit{$t \neq s, t \succ _F s$ if $\forall f \in F, f(t) \leq f(s)$\footnote{In the convention used here, lower values are better than higher ones.}}}\end{center} 

Where \textit{t F-dominates s}. It is shown in the same paper how \textit{F-dominance} is preserved considering also smaller sets of monotone functions, but it is not obvious applying it to larger ones and in \cite{ciacciaFDom} is also shown, using an example, how the concept can be applied to an infinite set of functions. The importance behind this formalization resides in the possibility of translating the user's preference in a mathematical constraint using a certain set of functions, leading to the ability to return an output that better satisfies the user himself.\\
Along with this definition, two R-skyline operators are presented in the same paper: \textbf{ND}, to characterize the set of non-\textit{F}-dominated tuples, and \textbf{PO}, that refers to the tuples that are potentially optimal, so the best ones according to a given function in \textit{F}.

The \textit{non-dominated restricted skyline} of \textit{r} w.r.t. \textit{F} is so defined as the set of tuples:
\begin{center} \large{ND(\textit{r;F}) = \textit{\{$t \in r \mid \not \exists s \in r, s \succ _F t$\}}}\end{center}

While the \textit{potentially optimal restricted skyline} of \textit{r} w.r.t. \textit{F} is defined as: 
\begin{center} \large{PO(\textit{r;F}) = \textit{\{$t \in r \mid \exists f \in F, \forall s \in r, s \neq t \Rightarrow f(t)<f(s) $\}}}\end{center}

And, in the general case, the following containment relation can be proven:
\begin{center}\large{PO(\textit{r;F}) $\subseteq$ ND(\textit{r;F}) $\subseteq$ SKY(\textit{r}).}\end{center}  
Combining the two operators with the \textit{F-dominance} concept, it is possible to identify a way to smoothly move from a skyline query style to a top-1 using the appropriate set of constraints or considering a given subset of the monotone functions.\\
Considering the set \textit{F} as coinciding with the whole monotone functions set \textbf{MF}, both the ND and the PO operators' output will be equal to the standard skyline query one. For an analogy with ranking queries instead, it can be observed that defining the \textit{F} family similarly to the scoring function used for the ranking query, a similar pattern in the results will be returned, taking however into account that the two operators will always provide a more interesting consideration on domination and potential optimality for the objects in question.\\ \\
For an example of application, consider a user browsing an online reviewer for internet pricing given different companies:

\begin{center}
\begin{tabular}{|c|c|c|}
\hline
Name & First semester payment & Yearly growth ratio \\
\hline
Accoli & 200 & 0.30 \\
Barenzi & 180 & 0.25 \\
Ciccoli & 200 & 0.15 \\
Danzi & 350 & 0.1\\
Effigi & 400 & 0.05\\
\hline
\end{tabular}
\end{center}

Applying a standard skyline query, Barenzi, Ciccoli, Danzi, and Effigi would be marked as more convenient w.r.t. Accoli. If we take into account the family of functions defined as: 
\begin{center}
\textit{F = \{$w_p FirstPayment + w_g YearlyGrowthRatio \mid w_p \geq w_g$}\}
\end{center}
Here, the user is giving more importance to initial price than yearly growth, other solutions can be cut out from the picture: in this way, Barenzi and Ciccoli would belong to the ND restricted skyline, while only Ciccoli would result as a Potentially Optimal solution.






\subsection{Trade-off Skyline}\label{troff}
This paradigm focuses on giving the concept of \textit{trade-off} in computing the given skyline, intended as a user decision between a couple of sample objects focusing on a subset of the available attributes \cite{tradeoffLofi, tradeoffLofiComp}. It is introduced to overcome the need of specifying in detail the weights needed to compute a scoring function, without having to deal with the possible high number of elements returned in a skyline, while allowing the possibility of a feedback factor from the user. \\ \\
First of all, the introduction of a new relationship is needed to implement the concept of \textit{trade-off}, starting from the assumptions that the two parameters on which the trade-off is carried out are incomparable with each other, as in the opposite case a trade-off wouldn't be necessary at all to define which choice was better. This notion can be defined as the qualitative description of how much the user is willing to renounce for convenience in some dimension(s) of the considered object, to gain better performance in some other dimension(s), starting from a practical example.

If a customer in an electronics shop states that he would pay even 15€ more for a cellphone to have it colored in black, considering two available products defined by color and price as $o_1=\{white, 10$\texteuro$\}$ and $o_2=\{black,20$\texteuro$\}$, using the new defined \textit{trade-off} relationship it would be possible to write:
\begin{center}\large{$t_1 := o_2 \triangleright o_1$}.\end{center}
Where $t_1$ is the trade-off just mentioned. Formally, a trade-off will always be defined over a set $\mu$ of attributes given by their respective indices, which is individual for every trade-off (i.e. different trade-offs can be defined on different sets). A new concept of domination, Trade-off domination, is so derived from this definition that allows extending domination to previously incomparable objects over a set of trade-offs: 
\begin{center}\large{The \textit{strict pre-order}\footnote{Binary relation that is irreflexive, transitive, and asymmetric.} $\succ_T$ is the smallest preorder containing the full product order\footnote{Order used to determine which elements dominate others in the database.} enhanced by additional domination relationships induced by trade-offs or arbitrary combinations of them in \textit{T}, set containing such combination.}\end{center}
Naturally, this domination concept can be applied to a skyline to obtain a \textit{trade-off skyline}. Considering $o_1$ and $o_2$ objects contained in database \textit{R}:
\begin{center}\large{$Sky := \{ o_2 \in R\mid \not \exists o_1 : o_1 \succ_T o_2\}$}.\end{center}
To define the same concept considering different pairs or combinations of these new relationships requires the use of more specific operators (e.g. in \cite{tradeoffLofiComp} the \textit{merge} operator) and deeper considerations that can be found in the referenced papers and in \cite{tradeoff1, tradeoff2}.\\ \\
Returning to the electronics shop example, if the shopkeeper would've used a storage system implementing this paradigm, he would've been able to return easily the number of objects which could dominate the others already present in the skyline given the trade-off expressed by the customer.

\subsection{Applications of $\rho$-dominance: ORD and ORU}\label{ORD}
\textbf{ORD} and \textbf{ORU} are the names of the two operators proposed in \cite{ORD}, capable of combining the main features of top-k and skyline ranking query. To better explain how should they suffice for this task, three hard requirements to meet in practical decision support are presented in the same paper: \textit{personalization}, \textit{control over output size} and \textit{flexibility in preference specification}. 

A base consideration is made on the available options, representing them as \textit{d}-dimensional records\\\textbf{\textit{r}} = $<x_1,x_2,...,x_d>$ in a \textit{D} dataset indexed with a spatial access method; in this context, the \textit{utility score} relative to a \textit{d}-dimensional preference vector of non-negative weights can be computed as their inner product:
\begin{center} \large{ \textit{$U_v(r) = \Sigma_{i=1}^d x_i \cdot w_i$}.}\end{center}
As specified in the same article, the ordering of \textit{D} by utility is independent of the order of magnitude of \textit{v}, which lets us assume preference vectors such that their inner sum of weights is equal to one: considering this fact from a geometrical point of view, it can be assessed that the \textit{preference domain} is equivalent to the (d-1)-simplex in a space where the axes correspond to the considered d-weights. Formally:
\begin{center} \large{\textit{$ \Delta^{d-1} = \{v \in \mathbb{R} \mid \Sigma_{i=1}^d w_i = 1\}$}.}\end{center}
Every valid preference is so represented as a vertex of this polytope.\\

With these premises, the concept of domination is also considered in a geometrical manner starting from the concept of \textit{$\rho$-domination}. Being \textit{w} the \textit{seed} (i.e. best effort approximation of user's preferences vector), and given different preference vectors \textit{v} with distance \textit{$\rho$} from \textit{w}, if a record \textit{$r_i$} scores at least as high as another record \textit{$r_j$} in all vectors and striclty higher for at least one of them, \textit{$r_i$ $\rho$-dominates $r_j$}. The records that are dominated by few in such way form the \textit{$\rho$-skyband}.

Using a different approach than the one seen in \ref{res}, using the restrictions on dimensions even in this case could be possible to develop a flexible solution to consider the passage from top-k to skyline queries, and the two introduced operators give an idea on the extent of such observation, stressing how the output size can be controllable in using them.

\begin{center} \large{Given the \textit{seed vector w} and the required output size \textit{m}, ORD will report the records that are \textit{$\rho$-dominated} by less than \textit{k} others, using the minimum \textit{$\rho$} that produces exactly \textit{m} results.}\end{center}

As a note, the appropriate \textit{$\rho$} is automatically calculated by the framework for its application in the definition, according to \textit{m}, as it happens for the second presented operator.

\begin{center} \large{Given the \textit{seed vector w} and the required output size \textit{m}, ORU will report the records belonging to the \textit{\textbf{top-k}} result for at least one preference vector within distance \textit{$\rho$} from \textit{w}, for the minimum distance that produces \textit{m} records in output.}\end{center}

For what concerns the experimental applications, in \cite{ORD} similar premises as the ones seen in \ref{troff} about the loss of meaning for multi-objective queries for high dimensions are done. The datasets used are both real and synthetic to observe how the computation performance for the algorithms associated with such operators may vary, and a comparison between them and other approaches is fully documented showing scalable performance among other aspects.


\subsection{Top-K dominating queries on skyline groups}\label{topkdom}
Skyline computation is very useful in multi-criteria decision-making applications but can result inadequate when applied to queries that need to focus on \textit{combinations} of these points, for when it comes to considering a finite group of points that one wants to extrapolate from given objects the user would be left performing this selection on an output that can be extremely vast compared with what he needed.

The solution proposed in \cite{topKDominatingZhu}, based on \cite{topKYiu} and \cite{progressiveSkyline}, tries to address this matter with the introduction of an algorithm to find the most representative k-skyline groups, combining the top-k and the skyline group approaches. To do so, some appropriate considerations a priori are needed:
\begin{itemize}
\item The definition of dominance is extended to groups. Given a point \textit{Q} and a group \textit{G}, \textit{Q} is dominated by \textit{G} \textit{iff} there exists at least one point \textit{Q'} in \textit{g} such that \textit{Q'}$\succ$\textit{Q}. This is given as the consequence of different dominance notions already introduced that are explained in the same paper;
\item For a given size \textit{l}, a group of \textit{l} points is a Skyline Group, or an \textit{l}-point Skyline Group, \textit{iff} is not dominated by any other \textit{l}-point group in the domain of interest \textit{D}. Skyline Groups (\textit{GSkylines}) is a set of all \textit{l}-points skyline groups;
\item A function \textit{score(G)} is defined as the relation returning the number of points dominated by group \textit{G}.
\end{itemize}

From these premises, Top-K Dominating query (\textbf{TKD}) on Skyline groups will return the set 

\begin{center}\large{\textit{$S_k\subseteq$ GSkylines}}\end{center} 

Of k Skyline groups with highest scores. So, to find the top-k dominating skyline groups, it will be necessary to check:

\begin{center}\large{\textit{$\forall G \in S_k, \forall G'$ in $(GSkyline - S_k) \Rightarrow score(G) \geq score(G')$}}\end{center} 

The result produced by this method is of a different kind w.r.t. the others presented in this survey, as it focuses its application on searching for a \textit{combination} of points that, under circumstances defined by some other operations, could result as dominating on the other groups defined; however, considering an output of this type, the application of these concepts for the resolution of a problem of similar formulation w.r.t. the other sections' cases is worthy of mention.\\ \\
For an example of application, take into account an online Dungeons\&Dragons character sheet compiler. Starting from a certain level, every player has to set an initial score for their character base stats, and every available class needs some higher characteristics instead of others to perform better; at the same time, the various points to apply in the sheet are bought from a finite pool. Some players, to have a performant character, could decide so to use an optimizer implementing this method to get the best (k=1) possible group w.r.t. a function expressing their needs in terms of stats.\\ \\
Given the general notion, different algorithms are introduced in the same paper w.r.t. the different operations under which TKD on skyline groups can be defined, with a consequent analysis of the time complexity of each.
As discussed in the article, the considerations made on the \textit{GSkylines} can improve the performance of pre-existing algorithms, but the cost of score computation can be considered too expensive as it is needed to check points dominated by others in a group multiple times so, in that same work, a possible solution is found in a bitmap indexing useful to compute such scores: to maintain an efficient computation, a bit vector is assigned for each point in a given group, to employ bitwise operations for much faster computation.
\\



\subsection{Uncertain top-k queries}\label{utk}
The strongest point of top-k ranking queries resides in the possibility for the user to personalize his choice using a certain scoring function, as introduced in \ref{topk}, defined over a certain set of weights-per-attribute; however, in practice, the desired preferences can be stated only with bounded accuracy or may be, in the worst case, uncertain, as it would be harsh or unfair for the user to establish with crucial precision an estimation on their preferences in some instances. To answer the need for a paradigm covering this scenario, \textit{Uncertain Top-k} queries were proposed in \cite{uncTopK}.\\
While other works dealt with uncertainty at the data level, as stated in the same paper, \textit{UTK} has the purpose of addressing the problem of uncertainty in the \textit{weight vector}.\\

The input of \textit{UTK} consists of a dataset \textit{D}, a positive integer \textit{k} and a \textit{region R}. Considering every record composed of \textit{d} values, it can be said that these attributes define a \textit{d}-dimensions data domain in which records can be seen as vectors. With premises on the score function similar to what has been discussed in \ref{ORD}, even in this case \textit{D} can be reduced to a (\textit{d}-1)-dimensional space, called \textit{preference domain}, in which the (\textit{d}-1)-dimensional form of the weight vector will be referred to as \textit{w} in subsequent definitions.
In \cite{uncTopK} a region is defined as an axis-parallel hyper-rectangle\footnote{A generalization of a rectangle on higher dimensions.} for ease of visualization but can be extended to a general convex polytope.

Considering \textit{R} in \textit{preference domain}, there is a distinction between two different UTK versions: 
\begin{center} \large{\textit{$UTK_1$} reports the \textit{exact} set of those records that may be included in the top-k when \textit{w} lies in \textit{R}, while \textit{$UTK_2$} returns the \textit{exact} top-k set for every possible \textit{w} in \textit{R}, partitioning the region in different zones in which the minimal set differs w.r.t. the position of \textit{w}.}\end{center}
For exact, it is intended the \textit{minimal} set, i.e. for every record \textit{p} in this set there is at least one weight vector \textit{w} in region \textit{R} for which that \textit{p} is included in the top-k set. In practice, using $UTK_1$ one would be able to establish which records among the ones present in the dataset are potentially among the top-k for weights in \textit{w} in region \textit{R}, while using $UTK_2$ the user would visualize a partitioning of \textit{R}, with the associated top-k records for each single partition, dependent to the value of \textit{w} in \textit{R}.\\ 

Considering such operators, a correlation can be seen between them and the framework described in \ref{res}. For the case $k=1$ the analogy between the two can be observed through the concept of \textit{convex hull}\footnote{The smallest convex polytope that encloses all
records in a dataset.} as ND and PO, operating strictly in this case, give as output records that surely belong to a portion of that polytope: a property of the convex hull resides in the capability of containing all the top-1 records for any weight vector \textit{w}. Similar reasoning is behind UTK, with the main difference residing in the possible exploration of more values for \textit{k}. 

Practically speaking, if we consider the possibility of having a set of linear functions on which it can be defined an \textit{F-}dominance \ref{res} the analogy can be observed between such set and the scoring function used to determine the best-performing objects that are returned by UTK, which assures the presence of the PO elements found with Restricted skylines in the output of the UTK operators for k=1.\\ \\
For a practical example, consider the Internet Companies table seen in \ref{res}, where the online portal has implemented an UTK methodology to support the user's searches, given a rectangular restricted region defined with $w_p \in [0.1;0.5]$ and $w_g \in [0.1;0.5]$ in which the weights vector can lie. Suppose the choice of considering not only the best option but even the $2^nd$ best (k=2): $UTK_1$ will output all of the offers that could appear in the top-2 for any combination of the weights considered in R, while $UTK_2$ will return a partition of that rectangle described on the plane built with $w_p$ and $w_g$ as axes with the relative top-2 available for every zone delimited. If the user would've considered instead only the top choice (k=1), then the result would've been Ciccoli again in $UTK_1$ and a single zone R in which Ciccoli would be the answer for $UTK_2$.



\subsection{Skyline ordering}\label{skyord}
The focus of this approach is given on \textit{size constrained skyline queries}, as the problem of returning \textit{exactly k} objects on a \textit{d} dimensions skyline that, for the purposes of the study, will remain unknown a priori. It is emphasized in \cite{skylineOrd}, in facts, how could also happen that the number of objects of interest requested by the user could be \textit{higher} than the cardinality of the skyline itself. The approach proposed in the same paper is called \textit{Skyline ordering} over the conventional Skyline query.

An informal definition of size constrained skyline queries can be:
\begin{center} \large{\textit{$Q^scs_k (P)$} = a subset \textit{S} of \textit{P}, composed by \textit{k} points that are seen as \textit{good} in terms of user interest, with \textit{P} being the d-dimensional dataset composed of a number \textit{p} of choices of interest.}\end{center}
The focus shifts so on finding a correct and efficient way to determine which points can be classified as \textit{good} in this instance. To this extent the \textit{Skyline Order} of a set of \textit{d-dimensional of points P} is formalized as a sequence \textit{S = $<S_1,S_2,...,S_n>$} given:

\begin{center} \large{ \begin{enumerate}
\item \textit{$S_1$} the Skyline of \textit{P} (i.e. \textit{$S_P$});
\item \textit{$\forall i$, $1<i<n$, $S_i$} the Skyline of \textit{$P\setminus \bigcup_{j=1}^{i-1} S_j$};
\item $\bigcup_{i=1}^n$ \textit{$S_i$} = \textit{P}.
\end{enumerate}}
\end{center}

Being \textit{$S_i$} the \textit{Skyline order subset} or \textit{Skyline subset} and n the \textit{Skyline order lenght}, as number of subsets in \textit{S}. By costruction it can be observed how skyline subsets are incomparable with each other and that \textit{$S_i \cap S_j = \O$ for $ i\neq j$} and for convenience is defined formally the index \textit{i} of the subset \textit{$S_i$} for a point \textit{p} such as:
\begin{center} \large{ \textit{$p \in S_i \rightarrow p \in P$}.}\end{center}

Given such premises, in the same work is derived a series of lemmas from which the central property of Skyline ordering is inferred: all the choices that dominate given \textit{$p \in S_i$} can be found only in skyline subsets \textit{$S_j$, with $j<i$}, while no choice can dominate any other in the same partition or in a previous one. This concept allows also the usage of some set-wide maximization techniques on such partitions, also described in the same paper.	\\
In that article, which reading is advised for a deeper understanding of the subject, several algorithms to compute such partitions efficiently and apply this paradigm in various contexts are presented with relative analysis.

\subsection{Regret minimization}\label{kreg}
As previously observed in \ref{utk}, uncertainty in the insertion of values on which the scoring function is computed is all but an unrealistic possibility, and the main concept behind this paradigm efficiently addresses this matter avoiding asking a scoring function to the user at all. Moreover, it provides an efficient way to return given points of interest to the user without overwhelming him with a too-large output, like could happen in the application of standard skyline queries; these features are possible thanks to the definition of \textit{regret ratio} for a user given a returned result, and of minimization for its maximum.\\

To define and use the notion of regret, in \cite{regretNanongkai} are first considered some necessary formalizations. First of all, the \textit{utility function} considered is mapped as an \textit{$f: \mathbb{R}^d_+ \rightarrow \mathbb{R}_+$}, and it is formalized that the utility for a point \textit{p} given a function \textit{f} will be written as \textit{f(p)}.\\Starting from it, the \textit{gain} derived from a subset of points \textit{$S \subseteq D$} where \textit{D} is the whole domain, is defined as:
\begin{center} \large{$\textit{gain(S,f)}=max_{p \in S}$ \textit{f(p)}.}\end{center}

And consequently the concept of \textit{regret} and \textit{regret ratio} can be introduced:

\begin{center}\large{Regret \textit{$r_D(S,f) = gain(D,f) - gain(S,f)$} and regret ratio \textit{$rr_D(S,f) = \frac{r_D(S,f)}{gain(D,f)}$}.}\end{center}

As a note, it can be seen how the regret ratio is $\in [0;1]$, where a 0 corresponds to a very happy user while a 1 means that he's very unhappy about choices returned. The last operator needed to compute the \textit{regret minimizing sets} is the \textit{maximum regret ratio}:

\begin{center}\large{\textit{$rr_D(S,F) = sup_{f \in F}$  $rr_D(S,f) = sup_{f \in F} \frac{max_{p \in D} f(p) - max_{p \in S} f(p)}{max_{p \in D} f(p)}$}.}\end{center}With \textit{F} being the class of utility functions considered in the computation. The main point of these definitions is to give a threshold that can be used to tune the \textit{k} points to return the user, starting from any set \textit{D} of \textit{n d}-dimensional objects:

\begin{center} \large{\textit{$\forall D$  $\exists K$ set of \textit{k} points such that \textit{$K \subseteq D$}, $rr_D(K,F) \leq \rho(k,d,n,F)$}.}\end{center}

From these new introductions, it is inferred the "happiness" in percentile of a user that would make a query on that \textit{K} subset instead of the whole available dataset, finding a way to represent the whole \textit{D} using just a subset of tuples while maintaining effectiveness on querying. 
In the same article is then considered a study on scale invariance and stability of this model, with the computation of a theoretical upper bound that is independent of the database size.\\
In practice, let's consider three different users: Arturo, Cristina, and Daniela, all of them searching for an apartment to rent in Milan considering cost, distance from their workplace, and the presence of an elevator in the structure. The online booking application on which they're all browsing solution, implementing the regret minimization approach, will be able to provide them the \textit{r} choices that are assured to leave them satisfied with the highest possible probability given the whole housing pool, without even knowing what the weight on their preferences was.\\

An extension of this previous work is presented in \cite{kRegretChester}, where it is formulated a \textit{k}-relaxation of the regret minimizing sets to address the possibility of considering not only the best possible element but even the \textit{$k^th$}, as it is presented how could be interesting for the user to know about more choices among the best ones in certain cases. Given a weight vector w, a certain point \textit{p} $\in \textit{D}$, a subset \textit{K} of domain \textit{D} and considering a score function defined as:
\begin{center}\large{\textit{$score(p,w) = \Sigma^{d-1}_{i=0} p_i w_i$}.}\end{center}
It is proposed a generalization of \textit{regret ratio} and its consisting notions to a top-k concept.
\begin{itemize}
\item \textit{$kgain(R,w) = score(R^{k,w},w)$;\footnote{The first input in score stays for the k-ed ranked element in R over w.}}
\item \textit{k-regratio(R,w) $= \frac{max(0,kgain(D,w)-gain(R,w)}{kgain(D,w)}$;}
\item \textit{With L denoting all vectors in $[0;1]^d$, k-regratio(R)$=sup_{w \in L}$ k-regratio(R,w);}  
\end{itemize}
And so, the focus of the problem shifts to finding a k-regret minimizing set of order (i.e. size) \textit{r} on a dataset \textit{D}, to minimize worst case scenario:
\begin{center} \large{\textit{$R_{r,D} =$ $argmin_{R \subseteq D, \mid R \mid = r}$ k-regratio(R)}.}\end{center}

Returning to Arturo's situation, it will be possible for him to get in output not only the singular optimal pool of choices, that could maybe have been booked in the meanwhile, but even the $2^{nd}$, the $3^{rd}$ or the \textit{$k^{th}$} one to browse and consider based on the worst-case scenario for happiness considerations.


\section{Comparison with vanilla approaches and conclusions}
As thoroughly described in this survey, each of the methods presented aims to address one or more specific restrictions of the standard top-k or skyline query approaches, that have been presented in more instances by the various cited researchers. In this section, a brief analysis of the considered methods is presented, emphasizing their common and/or unique characteristics.

\subsection{Restricted Skyline, ND and PO}
The concept of \textit{F-}dominance presented allows a broader observation of the results that would've been normally returned by a skyline, even reducing them to refine the standard output from the inconvenient objects w.r.t. the family of functions considered. The possible application of this dominance concept to an infinite set of functions paves the road for more complex definitions on the score concept, while with given restrictions the two operators and their algorithms can achieve results similar to the ones described in other paradigms like \ref{utk}. In the cited papers are also presented some peculiar interactions that the two operators have when the family \textit{F} takes the form of an $L_p$ norm. The flexibility of passing from a ranking to a skyline query is at the base of this same survey, and of crucial importance and utility.\\ 
Problems with this approach could be tied to the lack of control on precise output size for the query, and their applicability only to the "best object" ranking scenario (k=1); to suffice this second consideration, in \cite{uncTopK} the two operators are extended to cover even the k-est ranking as $ND_k$ and $PO_k$ \cite{ciacciaFDom}.
\subsection{Trade-off Skyline}
The only operator among the presented that provides a direct qualitative standard of comparison from a user's feedback, thanks to the formalization of the concept of \textit{trade-off}. It makes possible the comparison between objects that would've been incomparable otherwise, thanks to the novel concept of \textit{T-}domination, and in cases in which combinations of trade-offs are considered, in the same paper are presented methods and operators to correctly compute the skyline.\\ 
Its objective of improving Quality Of Life for users in certain instances is addressed in its functioning, but, as later specified by other paradigms introduced in this survey, it's not always possible for a customer to clearly state what his precise preferences would be given certain categories, especially when these grow too much in number. Moreover, computing the skyline is far from trivial in a combination of trade-offs, and requires the application of algorithms that can be expensive.
\subsection{Applications of $\rho$-dominance: ORD and ORU}
Both of the operators presented are highly focused on addressing the problem of uncertainty on output size using $\rho$-dominance, and their purpose is fulfilled as described by the papers cited, also taking into consideration personalization and flexibility for the user's preferences. As seen for R-skylines, even here it is possible to change the base structure of the query from ranking to skyline, introducing some restrictions on dimensions considered in computing the $\rho$ distance.\\
Even if personalization is one of the main pillars on which the authors pose the bases for their operators, the seed vector considered is an approximation of the user's preferences that should always be precisely expressed. Oh higher dimensions, the results returned begin to asymptotically grow nearer to the full domain even in this case. The appropriate $\rho$ needs to be specifically calculated by the framework from time to time to use correctly ORD and ORU, adding a certain computational time to the algorithms' functioning.
\subsection{Top-K dominating queries on skyline groups}
The method in question refers to an application that differs from the other observed, strictly addressing the problem of considering notions of dominance for combinations of objects instead of singular ones. As a hybrid between a top-k application and a skyline groups one, it presents a way to compute such ranking approaches given different operations on which it can be defined, using examples for each category of functions.\\
The result produced is strictly tied to the specific application case, and can't be extended to individual queries efficiently as the scoring function considered to define the concept of dominance applies the standard dominance concept to groups of objects, resulting in the application of a normal skyline query for singleton groups. It is needed a bitmap indexing strategy to maintain an efficient computation of the scoring function for groups, as without using bitwise operations it could be hardly useful as seen in \ref{topkdom}.
\subsection{Uncertain top-k queries}
One of the two methods presented to address uncertainty in weights considered to compute the scoring functions, and the only one between the two to give a precise indication of which records would appear in the query output for given weights considered. The usage of a convex polytope to determine top values in a multidimensional array is common to what is seen in other works, like \ref{res}, having as purpose the computation of the \textit{exact} set of records. $UTK_2$ gives an interesting visual display of how the output varies w.r.t. the value of the weight vector, and both the operators are flexible in determining even records that are close to the best one.\\
All the computations are made possible by a restriction of the preference domain to a specific region, which ensures again the best performance for the operators on low-dimensions. The option of taking into consideration the user's feedback in the computation that was made possible in other works, like \ref{troff}, here is ignored in favor of the possibility of returning top results for different weights considered. To compute such results is sometimes needed a higher complexity than what can be observed in similar output returning paradigms, like \ref{res} \cite{ciacciaFDom}.
\subsection{Skyline ordering}
The focus, like in \ref{ORD}, is posed on the need of returning exactly k objects as output in a skyline query. Instead of trying to introduce a novel domination variant, using a sequentialization of the space in which records can be found, it is ensured the construction of an output that can be contained in a smaller group; to apply this reasoning, it is necessary to define an ordering on the considered subsets.\\
The operations defined are useful for working on the subsets considered but have no impact at all on the objects contained inside the dataset, which will still be filtered through the standard concept of domination to compute the skyline. This method lacks a way to allow the user to express his preferences in terms of similar objects returned in the same partition, as the only input on which he's in control regards the size of the returned output, and the flexibility of considering best-performing entries instead of regular members of the starting skyline; both features are instead present in \ref{res}, \ref{utk} and, partially, in \ref{ORD}.
\subsection{Regret minimization}
The paradigm focuses on the concept of regret, as of unhappiness for the user, and exploits it to compute the combination of the solutions that would leave the whole population as happy as possible, without even knowing their initial preferences. The "probabilistic happiness-bound" approach adopted frees the user from the need of specifying the weights needed to compute the scoring function and allows him the practical visualization of the optimal solution in whatever case. Using this paradigm it is possible to identify a restricted domain in the whole dataset in which the usage of further operators will lead to a potential optimal result with more probabilities.\\
As seen for \ref{res}, the case of application was strictly limited to finding the best possible option, so even in this case in \cite{kRegretChester} is presented an extension of the concept of regret to k-regret that leads to a loss in terms of efficiency for cases not tied to k=1. Even considering the best possible outcome in this framework, there won't be an effective certainty for the final user on the optimality of returned result for him, only a good approximation of it.\\ \\ 
\subsection{Conclusions}
In this survey the basics for understanding some flexible solutions have been presented, born to overcome traditional ranking queries' inadequacies, each of them addressed in a peculiar, specific, and different way by one or more of them. For a deeper understanding of the subjects, it is always advised to read the appropriate sources, fully available on the web.

\bibliographystyle{plain}
\bibliography{SurveyFinal.bib}

\begin{thebibliography}{10}

\bibitem{tradeoff1}
Wolf{-}Tilo Balke, Ulrich G{\"{u}}ntzer, and Christoph Lofi.
\newblock Incremental trade-off management for preference-based queries.
\newblock {\em Int. J. Comput. Sci. Appl.}, 4(2):75--91, 2007.

\bibitem{kRegretChester}
Sean Chester, Alex Thomo, S.~Venkatesh, and Sue Whitesides.
\newblock Computing k-regret minimizing sets.
\newblock {\em Proc. {VLDB} Endow.}, 7(5):389--400, 2014.

\bibitem{restrictedMarti}
Paolo Ciaccia and Davide Martinenghi.
\newblock Reconciling skyline and ranking queries.
\newblock {\em Proc. {VLDB} Endow.}, 10(11):1454--1465, 2017.

\bibitem{ciacciaFDom}
Paolo Ciaccia and Davide Martinenghi.
\newblock Flexible skylines: Dominance for arbitrary sets of monotone
  functions.
\newblock {\em {ACM} Trans. Database Syst.}, 45(4):18:1--18:45, 2020.

\bibitem{topKQueryIlyas}
Ihab~F. Ilyas, George Beskales, and Mohamed~A. Soliman.
\newblock A survey of top-k query processing techniques in relational database
  systems.
\newblock {\em {ACM} Comput. Surv.}, 40(4):11:1--11:58, 2008.

\bibitem{kalyvas2017survey}
Christos Kalyvas and Theodoros Tzouramanis.
\newblock A survey of skyline query processing.
\newblock {\em CoRR}, abs/1704.01788, 2017.

\bibitem{kMostLin}
Xuemin Lin, Yidong Yuan, Qing Zhang, and Ying Zhang.
\newblock Selecting stars: The k most representative skyline operator.
\newblock In Rada Chirkova, Asuman Dogac, M.~Tamer {\"{O}}zsu, and Timos~K.
  Sellis, editors, {\em Proceedings of the 23rd International Conference on
  Data Engineering, {ICDE} 2007, The Marmara Hotel, Istanbul, Turkey, April
  15-20, 2007}, pages 86--95. {IEEE} Computer Society, 2007.

\bibitem{tradeoff2}
Christoph Lofi, Wolf{-}Tilo Balke, and Ulrich G{\"{u}}ntzer.
\newblock Efficiently performing consistency checks for multi-dimensional
  preference trade-offs.
\newblock In Oscar Pastor, Andr{\'{e}} Flory, and Jean{-}Louis Cavarero,
  editors, {\em Proceedings of the {IEEE} International Conference on Research
  Challenges in Information Science, {RCIS} 2008, Marrakech, Morocco, June 3-6,
  2008}, pages 271--278. {IEEE}, 2008.

\bibitem{tradeoffLofi}
Christoph Lofi, Wolf{-}Tilo Balke, and Ulrich G{\"{u}}ntzer.
\newblock Efficient skyline refinement using trade-offs.
\newblock In Andr{\'{e}} Flory and Martine Collard, editors, {\em Proceedings
  of the Third {IEEE} International Conference on Research Challenges in
  Information Science, {RCIS} 2009, F{\`{e}}s, Morocco, 22-24 April 2009},
  pages 353--364. {IEEE}, 2009.

\bibitem{tradeoffLofiComp}
Christoph Lofi, Ulrich G{\"{u}}ntzer, and Wolf{-}Tilo Balke.
\newblock Efficient computation of trade-off skylines.
\newblock In Ioana Manolescu, Stefano Spaccapietra, Jens Teubner, Masaru
  Kitsuregawa, Alain L{\'{e}}ger, Felix Naumann, Anastasia Ailamaki, and Fatma
  {\"{O}}zcan, editors, {\em {EDBT} 2010, 13th International Conference on
  Extending Database Technology, Lausanne, Switzerland, March 22-26, 2010,
  Proceedings}, volume 426 of {\em {ACM} International Conference Proceeding
  Series}, pages 597--608. {ACM}, 2010.

\bibitem{skylineOrd}
Hua Lu, Christian~S. Jensen, and Zhenjie Zhang.
\newblock Flexible and efficient resolution of skyline query size constraints.
\newblock {\em {IEEE} Trans. Knowl. Data Eng.}, 23(7):991--1005, 2011.

\bibitem{mindolin}
Denis Mindolin and Jan Chomicki.
\newblock Preference elicitation in prioritized skyline queries.
\newblock {\em CoRR}, abs/1008.5357, 2010.

\bibitem{ORD}
Kyriakos Mouratidis, Keming Li, and Bo~Tang.
\newblock Marrying top-k with skyline queries: Relaxing the preference input
  while producing output of controllable size.
\newblock In Guoliang Li, Zhanhuai Li, Stratos Idreos, and Divesh Srivastava,
  editors, {\em {SIGMOD} '21: International Conference on Management of Data,
  Virtual Event, China, June 20-25, 2021}, pages 1317--1330. {ACM}, 2021.

\bibitem{uncTopK}
Kyriakos Mouratidis and Bo~Tang.
\newblock Exact processing of uncertain top-k queries in multi-criteria
  settings.
\newblock {\em Proc. {VLDB} Endow.}, 11(8):866--879, 2018.

\bibitem{regretNanongkai}
Danupon Nanongkai, Atish~Das Sarma, Ashwin Lall, Richard~J. Lipton, and
  Jun~(Jim) Xu.
\newblock Regret-minimizing representative databases.
\newblock {\em Proc. {VLDB} Endow.}, 3(1):1114--1124, 2010.

\bibitem{progressiveSkyline}
Dimitris Papadias, Yufei Tao, Greg Fu, and Bernhard Seeger.
\newblock Progressive skyline computation in database systems.
\newblock {\em {ACM} Trans. Database Syst.}, 30(1):41--82, 2005.

\bibitem{introSkyline}
Eleftherios Tiakas, Apostolos~N. Papadopoulos, and Yannis Manolopoulos.
\newblock Skyline queries: An introduction.
\newblock In Nikolaos~G. Bourbakis, George~A. Tsihrintzis, and Maria Virvou,
  editors, {\em 6th International Conference on Information, Intelligence,
  Systems and Applications, {IISA} 2015, Corfu, Greece, July 6-8, 2015}, pages
  1--6. {IEEE}, 2015.

\bibitem{topKYiu}
Man~Lung Yiu and Nikos Mamoulis.
\newblock Efficient processing of top-k dominating queries on multi-dimensional
  data.
\newblock In Christoph Koch, Johannes Gehrke, Minos~N. Garofalakis, Divesh
  Srivastava, Karl Aberer, Anand Deshpande, Daniela Florescu, Chee~Yong Chan,
  Venkatesh Ganti, Carl{-}Christian Kanne, Wolfgang Klas, and Erich~J. Neuhold,
  editors, {\em Proceedings of the 33rd International Conference on Very Large
  Data Bases, University of Vienna, Austria, September 23-27, 2007}, pages
  483--494. {ACM}, 2007.

\bibitem{topKDominatingZhu}
Haoyang Zhu, Xiaoyong Li, Qiang Liu, and Zichen Xu.
\newblock Top-k dominating queries on skyline groups.
\newblock {\em {IEEE} Trans. Knowl. Data Eng.}, 32(7):1431--1444, 2020.

\end{thebibliography}
\end{document}